\documentclass[prd, aps, superscriptaddress, preprintnumbers, twocolumn, floatfix, nofootinbib]{revtex4}

\usepackage{amsfonts}
\usepackage{amsmath}
\usepackage{amssymb}
\usepackage{bm}
\usepackage{dcolumn}
\usepackage{graphicx}   
\usepackage[latin1]{inputenc}
\usepackage{latexsym}
\usepackage{rotating}
\usepackage{hyperref}
\usepackage{graphicx}
\usepackage{color}

\usepackage{amsfonts}
\usepackage{amsmath}
\usepackage{amssymb}
\usepackage{bm}
\usepackage{dcolumn}
\usepackage{graphicx}
\usepackage[latin1]{inputenc}
\usepackage{latexsym}
\usepackage{rotating}
\usepackage{hyperref}

\newcommand\be{\begin{equation}}
\newcommand\ba{\begin{eqnarray}}
\newcommand\ee{\end{equation}}
\newcommand\ea{\end{eqnarray}}

\begin{document}

\title {Tracking Dark Energy from Axion-Gauge Field Couplings}

\author{Stephon Alexander}
\email{stephonster@gmail.com}
\affiliation{Department of Physics, Brown University,
Providence, RI, 02912, USA} 

\author{Robert Brandenberger}
\email{rhb@physics.mcgill.ca}
\affiliation{Physics Department, McGill University, Montreal, QC, H3A 2T8, Canada, and \\Institute for Theoretical Studies,
ETH Z\"urich, CH-8092 Z\"urich, Switzerland}

\author{J\"urg Fr\"ohlich}
\email{juerg@phys.ethz.ch}
\affiliation{Institute of Theoretical Physics, ETH Z\"urich, CH-8093 Z\"urich, Switzerland}
\date{\today}

\begin{abstract}

We propose a toy model of Dark Energy in which the degrees of freedom currently dominating the 
energy density of the universe are described by a pseudo-scalar ``axion field'' linearly coupled to the Pontryagin density, 
$ \text{tr}(F \wedge F)$, i.e., the exterior derivative of the Chern-Simons form, of a gauge field. 
We assume that the axion has self-interactions corresponding to an exponential potential. 
We argue that a non-vanishing magnetic helicity of the gauge field leads to slow-rolling of the axion 
at field values far below the Planck scale. Our proposal suggests a ``Tracking Dark Energy Scenario'' in 
which the contribution of the axion energy density to the total energy density is constant 
(and small), during the early radiation phase, until a secular growth term proportional to the 
Pontryagin density of the gauge field becomes dominant. 
The initially small contribution of the axion field to the total energy density is related to the observed 
small baryon-to-entropy ratio. 
\end{abstract}

\pacs{98.80.Cq}
\maketitle

\section{Introduction} 

As is well known, the {\it Strong CP Problem} related to the vacuum structure of QCD, 
as described by the vacuum angle $\theta$, can be solved by promoting $\theta$ to a pseudo-scalar field, 
the \textit{axion} \cite{axion}. This field gives rise to a new species of light particles. 
It can be interpreted as the \textit{phase}, related to a $U(1)$- symmetry, of a complex 
scalar field \cite{PQ}. A non-vanishing vacuum expectation value of the scalar field 
leads to the spontaneous breaking of this symmetry. The field quanta of the axion are the 
Goldstone bosons accompanying the 
breaking of the $U(1)$- symmetry. They may acquire a mass through instanton effects and can be 
made ``invisible'' by choosing the symmetry breaking scale to be sufficiently high \cite{invisible}.
The axion may then be a candidate for {\it dark matter}; (see e.g. \cite{axionDM}). 
Some time ago, it has been suggested \cite{PF}
that, besides the QCD axion, there could exist an effective axion field conjugate to the anomalous 
axial vector current in QED. The time derivative of this axion field would then play the 
role of a space-time dependent chemical potential for the axial charge density in QED and, 
through the chiral anomaly, would give rise to an instability triggering the growth of 
low-frequency magnetic fields with non-trivial helicity; see also \cite{W}.
Possible applications of this observation to early universe cosmology, and in particular to the issue of the
generation of primordial magnetic fields, have been discussed in \cite{BFR}; (see also \cite{JS}, \cite {W}).

In this paper, we explore the possibility that an axion field, $\phi$, linearly coupled to the Pontryagin- or ``instanton'' density,
$\text{tr}(F\wedge F)$, of a non-abelian gauge field, $A_{\mu}$, could contribute to the {\it dark energy} of the Universe. 
We assume that the axion field has non-trivial self-interactions described by a potential term, $V(\phi)$, in the action functional. 
As has previously been observed in the context of inflationary models in \cite{Sorbo, Wyman} (see also \cite{Shahin}), 
the coupling of the axion to the instanton density of $A_{\mu}$ can lead to {\it slow-rolling} of
$\phi$ also for values of $\phi$ much smaller than the Planck mass. We show that
slow-rolling of $\phi$ leads to a {\it ``tracking solution''} with the property that the energy density of $\phi$
tracks that of the radiation-dominated background of the early Universe up to a 
time $t_c$ when a secular growth term in the magnetic helicity of the gauge field starts to
dominate. From this time on, the contribution of $\phi$ to the energy density
of the Universe starts to grow until it might actually dominate it at some late
time. Choosing parameter values motivated by the observed small baryon-to-entropy
ratio, we arrive at a scenario in which the currently observed dark energy in the Universe may come from
 the axion field $\phi$. Thus, our mechanism might represent a realization of the {\it tracking dark energy} scenario previously 
discussed in \cite{tracking}; (see also \cite{Ratra}).

In the following section we describe some key features of our scenario. One
such feature is a secular growth in cosmological time of the electric component of 
the gauge field tensor. This is discussed in more detail in Section III, 
where we derive the gauge field equations of motion in the presence of a
term coupling the Pontryagin density, i.e., the exterior derivative of the Chern-Simons 3-form, 
to the axion field. We then attempt to find solutions of these equations that yield a homogeneous and isotropic 
Pontryagin density. It turns out that such solutions only exist for non-abelian gauge fields.
In Section IV we consider an exponential
potential for the self-interactions of $\phi$ and try to find out under what conditions tracking dark energy
arises.  In Section V we discuss tentative
particle physics connections of our scenario. Some conclusions are presented in Section VI. 
An interesting variant of our scenario involving a complex scalar field whose phase plays the role 
of the new axion field introduced in the present paper will be discussed in forthcoming work.
 
A word on our notation: Our space-time metric has signature
$(-,+,+,+)$. We work in units in which the speed of light, Planck's constant
and Boltzmann's constant are all set to $1$. The cosmological scale factor
is denoted by $a(t)$, where $t$ is physical time. The Hubble expansion rate
is $H(t) = \frac{\dot a}{a} (t)$, and $t_{eq}$ denotes the time of equal matter
and radiation.

\section{Key Features of Our Scenario} 

In this section we introduce our dark energy model, postponing a discussion of 
its origins in particle physics to Section V.

A key element of our model is a pseudo-scalar axion field, $\phi$,
that couples linearly to the Pontryagin density, $\text{tr}(F \wedge F)$, of a (massive) non-abelian gauge field, $A_{\mu}$.
The dynamics of the axion field $\phi$, the gauge field $A_{\mu}$ and the space-time metric 
$g_{\mu \nu}$ is determined by the following action functional:
\be
S  \, = \, \int {d^4x} \sqrt{-g}  \left[ \frac{R}{16\pi G} + {\cal L}_m \right] \, ,
\ee
where the matter Lagrangian is given by 
\ba
{\cal L}_m \, &=& \, \frac{1}{2} \partial_\mu \phi\partial^\mu\phi - V(\phi)   \\
& & -\frac{1}{4}F_{a\mu\nu}F_{a}^{\mu\nu}  - \frac{\lambda}{f} \phi F_{a\mu\nu}{\tilde{F}}_{a}^{\mu\nu} + \text{mass terms}\, . \nonumber
\ea
The second but last term, henceforth called \textit{`` magnetic helicity term''}, 
can be understood as arising from coupling the gradient of $\phi$ 
to an anomalous axial vector current and then invoking the chiral anomaly \cite{ABJ}. We will discuss 
possible particle physics origins of the field $\phi$, of an anomalous axial vector current, and 
of a heavy gauge field (with field strength denoted by $F$) in Section V.
In Eq. (2), repeated indices are to be summed over, the index $a$ is a gauge group index, $\mu$ and $\nu$ are space-time
indices, $\lambda$ is a dimensionless coupling constant, and $f$ is a reference field value that also 
appears in the axion potential $V(\phi)$.
In this paper we consider an exponential potential:
\begin{equation}
V(\phi) \, = \, \mu^4 e^{\phi / f} \, ,
\end{equation}
where  $\mu$ sets the energy
scale of the potential. This choice of $V(\phi)$ leads to an explicit breaking of parity and time-reversal invariance. 
(To avoid this, one might consider replacing $\text{exp}(\phi/f)$ by $\text{cosh}(\phi/f)^{-1}$ in Eq. (3).) 
A more natural choice of self-interactions not breaking these symmetries explicitly will be considered in forthcoming work.

A basic feature of our model is related to the expectation that
$\phi$ is very slowly rolling at sub-Planckian field values,
due to its coupling to the gauge field.
Assuming spatial homogeneity, the field equation of motion for $\phi$ is given by
\be \label{axionEq}
{\ddot \phi} + 3 H {\dot \phi} + V'(\phi) \, = \, \frac{\lambda}{8 f} {\vec{E}_{a}} \cdot {\vec{B}_{a}} \, ,
\ee
where the prime denotes a derivative of $V$ with respect to $\phi$. 
Following a hypothesis introduced in the context of inflationary models in \cite{Sorbo} and in
\cite{Wyman}, we assume that the term proportional to the Pontryagin is responsible for
slow-rolling of $\phi$, in the sense
that the terms in (4) proportional to first and second time derivatives of $\phi$
are negligible as compared to the two remaining terms. If this assumption can be justified
the equation of motion for $\phi$ reduces to 
\be \label{SReq1}
V'(\phi) \, \simeq \,  \frac{\lambda}{8 f} {\vec{E}_{a}} \cdot {\vec{B}_{a}} \, ,
\ee
an equation that determines the time-dependence of $\phi$,
once one knows the time-dependence of ${\vec{E}_{a}} \cdot {\vec{B}_{a}}$. We note
that slow rolling can arise at sub-Planckian field values, \cite{Sorbo},
in contrast to the usual slow-roll in large-field inflationary scenarios,
which requires super-Planckian values. In the context of inflation,
a scenario based on the two basic features of our model described so
far is sometimes called {\it chromo-natural} inflation \cite{Wyman}.

A third key feature of our scenario concerns the secular growth of a spatially homogeneous configuration 
of the electric field $E_{a}$, in excess of its usual dynamics. This growth is
induced by the coupling of the gauge field to the axion field $\phi$, as in (2). Under the assumption that 
the coupling constant $\lambda$ is sufficiently large, we will show that 
secular growth of $E_{a}$, when combined with Eq. (5), yields an axion field configuration 
that gives rise to {\it tracking dark energy}.

The main point is that a non-vanishing magnetic helicity, which originates
from the coupling of the gauge field to the axion as expressed by the ``magnetic helicity term,'' 
acts as an extra ``friction term'' that ensures that $\phi$ will slowly roll down its potential --
even at sub-Planckian field values. Thus, the resulting equation of state for the 
energy density of the axion is dominated by
the potential energy term, which yields a contribution
to dark energy. Equations (3) and (5) then tell us that if the Pontryagin density 
$\vec{E}_{a}\cdot \vec{B}_{a}$ exhibits secular growth, the contribution of $\phi$ 
to the total energy density of the Universe can become important at late times.

Finally, another key feature of our scenario is that the
\textit{initial value} of the energy density of $\phi$ is proportional to a
small number in cosmology, such as the baryon to entropy ratio $n_b/s$,
($n_b$ and $s$ being the baryon and photon number densities, respectively).
As far as relating a late-time cosmological observable to the
small baryon to entropy ratio (via a term in the Lagrangian
coupling the axion to an anomalous current) is concerned 
there are similarities of our work to the one in \cite{Stephon},
where the tensor-to-scalar ratio, (i.e., the ratio of the strength of
gravitational waves to that of scalar cosmological fluctuations),
is related to $n_b/s$.

\section{Gauge Field Dynamics in the Presence of the ``Anomaly Term''}

The equation of motion for the field strength tensor of the gauge field in the presence
of a Chern-Simons term (but neglecting mass terms) 
is given by
\begin{equation}
D^{ab}_\alpha F^{b\alpha \beta} -
\frac{4\lambda}{f}\epsilon^{\mu\nu\beta\alpha}\partial_\alpha^{ab}(\phi F_{\mu\nu}^b) = 0 \, .
\end{equation}
In this equation $D$ denotes the covariant derivative, which is defined by
\be
D_\alpha^{ab}\equiv \delta^{ab}\nabla_\alpha-gf^{abc}A^{c}_\alpha
\ee
where $\nabla_\alpha$ is the space-time covariant derivative, $g$ is the gauge coupling constant, and 
the $f^{abc}$ are the structure constants of the gauge group.

We write the equation of motion for the gauge field in terms of the 
``electric'' and ``magnetic'' components of the field tensor,
\ba
E^a_\mu \, &=& \, F^a_{\mu\nu}u^\nu, \\
B^a_\mu \, &=& \,  -\frac{1}{2}\epsilon_{\mu\nu\rho\sigma}F^{a\rho\sigma}u^\nu\, , \nonumber
\ea
where $u^\mu=(1,0,0,0)$ is the four-velocity of a comoving observer in an FRW spacetime.  In 
manifestly covariant form, the equations of motion are
\ba
u^\alpha D^{ab}_\alpha E^{b\sigma} &+& 
2HE^{a\sigma} - u_\mu\epsilon{^{\mu\sigma\alpha\beta}}D_\alpha^{ab}B_\beta^b \nonumber \\
&=& -\frac{8\lambda}{f}\left(u_\mu\epsilon{^{\mu\sigma\alpha\beta}}\partial_\alpha \phi E_{\beta}^a +u^\alpha\partial_\alpha\phi B^{a\sigma}\right), \nonumber \\
D^{ab}_\alpha E^{b\alpha} &=& \frac{8\lambda}{f}\partial_\alpha\phi B^{a\alpha} .
\ea
In standard three-vector form, the equations are
\ba
D_0^{ab}\mathbf{E}^b &+& 2H\mathbf{E}^a - \frac{1}{a}\mathbf{D}^{ab}\times \mathbf{B}^b \\
&=&  -\frac{8\lambda}{f}\left(\frac{1}{a}\nabla\phi\times\mathbf{E}^a +\dot{\phi}\mathbf{B}^a\right) \nonumber \\
\mathbf{D}^{ab}\cdot \mathbf{E}^b &=& \frac{8\lambda}{f}\nabla\phi\cdot\mathbf{B}
\ea
where $\mathbf{D}^{ab}$ is the spatial part of $D^{ab}_\alpha$.

Using the general definition of the electric and magnetic components of the field tensor, 
\ba
E_{ai} &=& \partial_{0}A_{i}^{a} -D^{ab}_{i}(A)A_{0}^{b} 
\ea
and
\ba
B_{ai} &=& \epsilon_{ijk}(\partial_{j}A_{k}^{a} -\frac{g}{2}\epsilon^{abc}A^{b}_{j}A^{c}_{k})
\ea
the field equations for the gauge field coupled to the axion field take the form
\ba 
\frac{\partial}{\partial t}E^{i}_{a} &-& \frac{g}{2} \epsilon^{abc}A_{0b}E^{i}_{c} +  2HE^{i}_{a} - \varepsilon^{ijk}\nabla_{j}B_{ak} \\
&=& -\frac{\lambda}{f}[\dot{\phi}B^{i}_{a} -\varepsilon^{ijk} \nabla_{j}\phi  E_{ak}] \nonumber
\ea
and
\be 
\nabla_{i}E^{i}_{a} = -\frac{\lambda}{f} \nabla_{i}\phi B^{i}_{a},
\ee
with
\be 
\varepsilon^{ijk}\nabla_{j}  E_{ak} = -\frac{\partial}{\partial t} B^{i}_{a} \, 
\ee

In the following we will consider a spatially homogeneous
gauge field configuration.
Of course, in general non-vanishing electric and magnetic background fields
break rotational invariance; but spatial homogeneity can be preserved in gauge-invariant 
combinations of $\vec{E}$ and $\vec{B}$. As a result of a non-vanishing instanton condensate, 
$CP$ is broken, which may have interesting consequences that we
will return to elsewhere. The instanton condensate $\vec{E}_{a} \cdot \vec{B}_{a}$ 
can be non-zero, homogeneous \textit{and} isotropic,
i.e., translation- \textit{and} rotation invariant. 

For a non-Abelian gauge group, such as $SU(2)$, we make the ansatz
 of a spatially homogeneous background gauge field. Following
\cite{Adshead2} (see also \cite{Armen}) we take
\ba
A_0 \, &=& \, 0 \\
A_i(t) \, &=& \, \ a \psi(t) \delta_i^a J_a \, \nonumber
\ea
where $J_a$ are the generators of the non-Abelian $SU(2)$ gauge group and
$\delta^{i}_{a}$ is a Kronecker delta symbol combining an upper
internal (Lie-algebra) index, $a$, with a lower spatial index, $i$.
The field strength tensor elements are then
\ba
F_{0i}^a \, &=& \, a {\dot{(a \psi)}} \delta_i^a \nonumber \\
F_{ij}^a \, &=& \, - g (a \psi)^2 \epsilon^a_{ij} \, ,
\ea
which in particular implies that the electric field
can be written as
\be \label{twenty}
E_{i}^{a} \, \sim \, E(t) \delta_{i}^{a} \, .
\ee

Before the gauge field acquires a mass, 
the equation of motion for $\psi$ is (see \cite{Wyman})
\be \label{psieq1}
{\ddot{\psi}} + 3 H {\dot{\psi}} + \bigl( {\dot H} + 2 H^2 \bigr) \psi + 2 {\tilde g}^2 \psi^3
\, = \, {\tilde g} \frac{\lambda}{f} {\dot{\phi}} a^{3/2} \psi^2 \, .
\ee
where the term on the right hand side of the equation comes
from the magnetic helicity (instanton) term in the action which leads
to a coupling of the axion to the background gauge field.
The coefficient of the term linear in $\psi$ vanishes in the
radiation epoch. 

The equation (\ref{psieq1}) displayed above holds at all times
for an unbroken gauge theory. However, we are more interested
in a gauge theory that is spontaneously broken at
a fairly large mass scale $m$. The gauge field of the theory acquires its
mass after the symmetry breaking phase transition, a
transition occurring when the temperature, $T(t)$, of the Universe 
is of the order of $m$.  After the phase transition, at temperatures below the transition temperature, 
the energy density of the gauge field 
scales like that of matter, i.e.,
$\rho_{gauge} \sim a(t)^{-3}$, for times greater than $t_m$, where
$t_m$ is determined by
\be
T(t_m) \, \approx \, m \, ,
\ee
where $T$ denotes the temperature of radiation.This corresponds to the scaling
\be
E(t), B(t) \, \sim \, a(t)^{-3/2} \, .
\ee

Once the gauge field
acquires a mass, there will be an extra mass term in
the equation of motion for $\psi$, yielding
\ba \label{psieq2}
{\ddot{\psi}} &+& 3 H {\dot{\psi}} + \bigl( {\dot H} + 2 H^2 \bigr) \psi 
+ m^2 \psi + 2 {\tilde g}^2 \psi^3 \nonumber \\
&=& \, {\tilde g} \frac{\lambda}{f} {\dot{\phi}} a^{3/2} \psi^2 \, .
\ea
Since we are assuming that the
mass will be much larger than the value of $H$ at
the time of equal matter and radiation, the third term on
the left hand side of (\ref{psieq2}) is negligible even
in the matter era. 
The nonlinear term proportional to $\psi^3$
becomes increasingly unimportant compared to
the ${\ddot \psi} \sim m^2 \psi$ term as time goes on since
the amplitude of $\psi$ decreases. 
The approximate form of (\ref{psieq2})
then becomes
\be \label{psieq3}
{\ddot{\psi}} + 3 H {\dot{\psi}} + m^2 \psi
\, = \, {\tilde g} \frac{\lambda}{f} {\dot{\phi}} a^{3/2} \psi^2 \, .
\ee
In terms of electric and magnetic fields, this equation
corresponds to (suppressing the gauge group index
$a$)
\begin{eqnarray} \label{EqoM3}
\frac{\partial}{\partial t}E_{i} + \frac{3}{2} H E_{i}\, 
&=& \, -\frac{\lambda}{f}[\dot{\phi} B_{i} ] \, \nonumber  \\
{\dot B}_{i} \, + \, \frac{3}{2} H B_{i} \, & = & \, 0 \, 
\end{eqnarray}

The key point is that the coupling of the gauge field to the axion field 
$\phi$ has the consequence that the electric field decays less rapidly 
than it would in the absence of $\phi$. This effect entails that the 
magnetic helicity and the energy density, 
\mbox{$\frac{1}{2}(\vec{E}^{2}+\vec{B}^{2})$} of the gauge field,
which, initially, scale like that of radiation, then like the one of matter, turn out to grow 
relative to the energy density of matter, at late times.

In the absence of the magnetic helicity term proportional to $\dot{\phi}$, or if the $\phi$-field is
time-independent, the equations (\ref{psieq3}) or (\ref{EqoM3}) imply the behavior 
$E_{i} \sim a^{-3/2}$ and $B_{i} \sim a^{-3/2}$. Hence they imply that the energy density
of the gauge field (i.e., of the $\vec{E}$- and $\vec{B}$- fields) scales like that of matter 
(For times $t < t_m$,
the energy density of the $\vec{E}$- and $\vec{B}$- fields decays like that of radiation.
This corresponds to a scaling of these fields proportional to $a^{-2}$.
Let us denote the $\psi$ field in the absence of the ``magnetic helicity term" 
by $\psi_0(t)$.

We may then determine the effects caused by the ``magnetic helicity term'' using
 the Green function method. We write
 \be
 \psi(t) \, = \, \psi_0(t) + \psi_1(t) \, ,
 \ee
 where $\psi_0(t)$ is the solution without the magnetic
 helicity term which satisfies the given initial conditions at an
 initial time $t_i$,
 and $\psi_1(t)$ is the first order Born approximation calculated from
\be \label{Born}
{\ddot{\psi_1}} + 3 H {\dot{\psi_1}}  + m^2 \psi_1
\, = \, {\tilde g} \frac{\lambda}{f} {\dot{\phi}} a^{3/2} \psi_0^2 \, \equiv \, S(t) \, ,
\ee
where $S(t)$ stands for the ``source'' term in the Born approximation. The
solution can be written as
\be \label{psiGF}
\psi_1(t) \, = \, \int_{t_i}^{t} dt' G(t, t') S(t') \, ,
\ee
where $G(t, t')$ is the Green function which is given by
\be \label{help1}
G(t, t') \, = \, w(t')^{-1} \bigl( u_1(t) u_2(t') - u_2(t) u_1(t') \bigr) \, ,
\ee
where $u_1$ and $u_2$ are the two basis solutions of the
homogeneous equation of motion, and
\be 
w(t) \, = \, {\dot u_1}(t) u_2(t) - {\dot u_2}(t) u_1(t)
\ee
is the Wronskian.

In the radiation epoch, i.e. for $t < t_{eq}$ we have
\ba \label{help2}
u_1(t) \, &=& \, \bigl( \frac{t_i}{t} \bigr)^{3/4} {\rm cos}(m t) \, \nonumber \\
u_2(t) \, &=& \, \bigl( \frac{t_i}{t} \bigr)^{3/4} {\rm sin}(m t) \, .
\ea
Hence the leading term in the Wronskian gives
\be \label{help3}
w(t') \, \simeq \, m \bigl( \frac{t_i}{t'} \bigr)^{3/2} \, .
\ee
As we will see later
\be \label{help4}
\frac{1}{f} |{\dot \phi}| \, = \, {\tilde n} \frac{1}{t} \, ,
\ee
where ${\tilde n}$ is a number of order one. Computing (\ref{psiGF}),
using \eqref{help1}, \eqref{help2}, \eqref{help3}, and \eqref{help4}, yields a term $\psi_0(t)$,
multiplied by a certain integral denoted $G(t)$,
\be
\psi_1(t) \, = \, \psi_0(t) G(t) \, .
\ee
Here $G(t)$ is a ``secular growth factor'' given by
\be \label{growth}
G(t) \, \simeq \, {\tilde g} {\tilde n} \lambda m^{-1}  \psi_0(t_i) {\rm log}(\frac{t}{t_i}) \, .
\ee
The secular growth of $\psi$ leads to a corresponding secular growth
of the magnetic helicity ${\rm tr} ({\vec{E}} \cdot {\vec{B}} )$. The analysis for
$t > t_{eq}$ is analogous, except that the mode functions $u_1$ and $u_2$
now scale as $t^{-1}$.

Based on the above analysis we are able to determine the scaling of
the magnetic helicity term. Taking into account the fact that
$B(t)$ scales as $a(t)^{-2}$, for $t < t_m$, and as $a(t)^{-3/2}$, for
$t > t_m$, we find that
\be
{\vec{E}_{a}} \cdot {\vec{B}_{a}} \, \sim \, a(t)^{-4}
\ee
for $t < t_m$, as
\be
{\vec{E}_{a}} \cdot {\vec{B}_{a}} \, \sim \, a(t)^{-3} 
\ee
for $t_m < t < t_c$, and as
\be
{\vec{E}_{a}} \cdot {\vec{B}_{a}} \, \sim \, a(t)^{-3} G(t)\, ,
\ee
for $t > t_c$.

\section{Late Time Acceleration for an Exponential Potential}

Next, we analyze how the dynamics of the axion $\phi$ is affected by gauge field configurations 
with non-vanishing magnetic helicity. 
We recall that, for a spatially homogeneous configuration of axions, the field equation of $\phi$ 
is given by
\be \label{axionEq}
{\ddot \phi} + 3 H {\dot \phi} + V'(\phi) \, = \, \frac{\lambda}{8 f} {\vec{E}_{a}} \cdot {\vec{B}_{a}} \, ,
\ee
where the prime indicates a derivative of $V$ with respect to $\phi$. 
As in the context of inflationary models in \cite{Sorbo}, we assume
that the term on the right side of \eqref{axionEq} generates slow-rolling of $\phi$, namely
that the terms $\ddot{\phi}$ and $3H\dot{\phi}$ in (\ref{axionEq})
are negligible as compared to the two remaining terms. We will check
the self-consistency of this assumption below. The evolution 
of $\phi$ is then determined by
\be \label{SReq1}
V'(\phi) \, = \,  \frac{\lambda}{8 f} {\vec{E}_{a}} \cdot {\vec{B}_{a}} \, 
\ee

For an exponential potential,
\be \label{poteq}
V(\phi) \, = \mu^4 e^{\phi / f},
\ee
Eq. (\ref{SReq1}) yields
\be \label{SReq2}
V(\phi) \, = \, \frac{\lambda}{8} {\vec{E}_{a}} \cdot {\vec{B}_{a}} \, 
\ee
The proportionality of $V(\phi)$ to $\text{tr }({\vec{E}} \cdot {\vec{B}})$ is a special
feature of the exponential potential. For polynomial and
periodic potentials, $V(\phi)$ ends up being proportional to a power of
$\text{tr }({\vec{E}} \cdot {\vec{B}})$ greater than $1$, and hence would
decay faster in time, in an expanding universe. This makes it more difficult
 to interpret $\phi$ as a dark-energy candidate. 
We will study such types of potentials in the context of a different (possibly more 
natural) model in a follow-up paper.

Combining (\ref{poteq}) and (\ref{SReq2}) we find the following
expression for the axion field $\phi$:
\be
\frac{\phi}{f} \, = \, {\rm log} \bigl( \frac{\lambda}{8} \mu^{-4} {\vec{E}_{a}} \cdot {\vec{B}_{a}} \bigr) \, ,
\ee
which leads to
\be \label{deriv}
\frac{1}{f} |{\dot{\phi}}| \, = \, {\tilde n} \frac{1}{t} \, ,
\ee
where the number ${\tilde n}$ is ${\tilde n} = 2$, for $t < t_m$ and for $t > t_{eq}$, and 
equals ${\tilde n} = 3/2$, for $t_m < t < t_{eq}$. This expression is
important for the evaluation of the magnitude of the secular growth term
 (\ref{growth}). In fact, inserting (\ref{deriv}) into (\ref{EqoM3}) and
taking into account that $B$ and $E_{0}$ scale the same way as a
function of time we find that
\be
E_{a}(t) \, = \, E_{a0}(t) \bigl[ 1 + {\tilde n} \lambda \frac{B_{a}(t_i)}{E_{a}(t_i)} {\rm log} (\frac{t}{t_i}) \bigr] \, .
\ee
Thus, the secular growth term grows logarithmically in time.

We now search for conditions implying that $\phi$ is a viable
candidate for {\it dark energy}. First of all, $\phi$ has to
be slowly rolling as a function of time in order for the
equation of state of $\phi$ to be that of dark energy. Second, we
have to show that the energy density of $\phi$ has the potential 
to dominate over the background energy density shortly
before the present time. To complete our analysis, we need to make sure that
the energy density of the gauge field which the axion field $\phi$ couples to remains
subdominant.

Under the assumption that the slow-rolling conditions are satisfied,
Eq. (\ref{SReq2}) immediately leads to an expression for the
contribution, $\Omega_{\phi}$, of the $\phi$- field to the total energy
density of the universe. From the above discussion it follows that, for
$t < t_m$, the energy density of $\phi$ scales like that of
radiation and hence leads to a constant contribution to
$\Omega_{\phi}$. For $t_m < t < t_{eq}$, the potential
energy of $\phi$ decreases less fast than the background
radiation density, leading to a contribution to $\Omega_{\phi}$
that grows linearly in $a(t)$. We should emphasize, however,
that the contribution of $\phi$ to $\Omega$ scales in the
same way as the contribution of dark matter to $\Omega$.
Once $t > t_{eq}$, but before
$t = t_c$, both the background density and the energy density
of $\phi$ scale as $a(t)^{-3}$, and hence the contribution of
$\phi$ to $\Omega$ is constant. Finally, once $t > t_c$,
 $\Omega_{\phi}$ increases in time.
Specifically, for late times $t > t_{eq}$, we obtain that
\ba \label{tracking}
\Omega_{\phi}(t) \, & \simeq & \, \frac{V(\phi(t))}{\rho_0(t)} \\
& = & \, \frac{\lambda}{8} \frac{({\vec{E}} \cdot {\vec{B}})(t_i)}{\rho_r(t_i)} \bigl( \frac{a(t_{eq})}{a(t_m)} \bigr)
\bigl[ 1 +   {\tilde n} \lambda \frac{B(t_i)}{E(t_i)} {\rm log}(\frac{t}{t_i}) \bigr] \, , \nonumber
\ea
where $\rho_0(t)$ is the background energy density at
time $t$, and $\rho_r(t_i)$ is the energy density of radiation at the initial time $t_i$, 
(which is approximately equal to the total energy density at that time, since we have
assumed that $t_i$ is chosen to belong to the radiation period). 
Equivalently, we can express $\Omega_{\phi}$ in terms of the background
matter density, $\rho_m$, at the initial time
\ba \label{tracking2}
\Omega_{\phi}(t) \, & \simeq & \, \frac{V(\phi(t))}{\rho_0(t)} \\
& = & \, \frac{\lambda}{8} \frac{({\vec{E}} \cdot {\vec{B}})(t_i)}{\rho_m(t_i)} \bigl( \frac{a(t_i)}{a(t_m)} \bigr)
\bigl[ 1 +   {\tilde n} \lambda \frac{B(t_i)}{E(t_i)} {\rm log}(\frac{t}{t_i}) \bigr] \, , \nonumber
\ea
Figure 1 presents a sketch of the time evolution of $\Omega_{\phi}$.

\begin{figure*}[t]
\begin{center}
\includegraphics[scale=0.6]{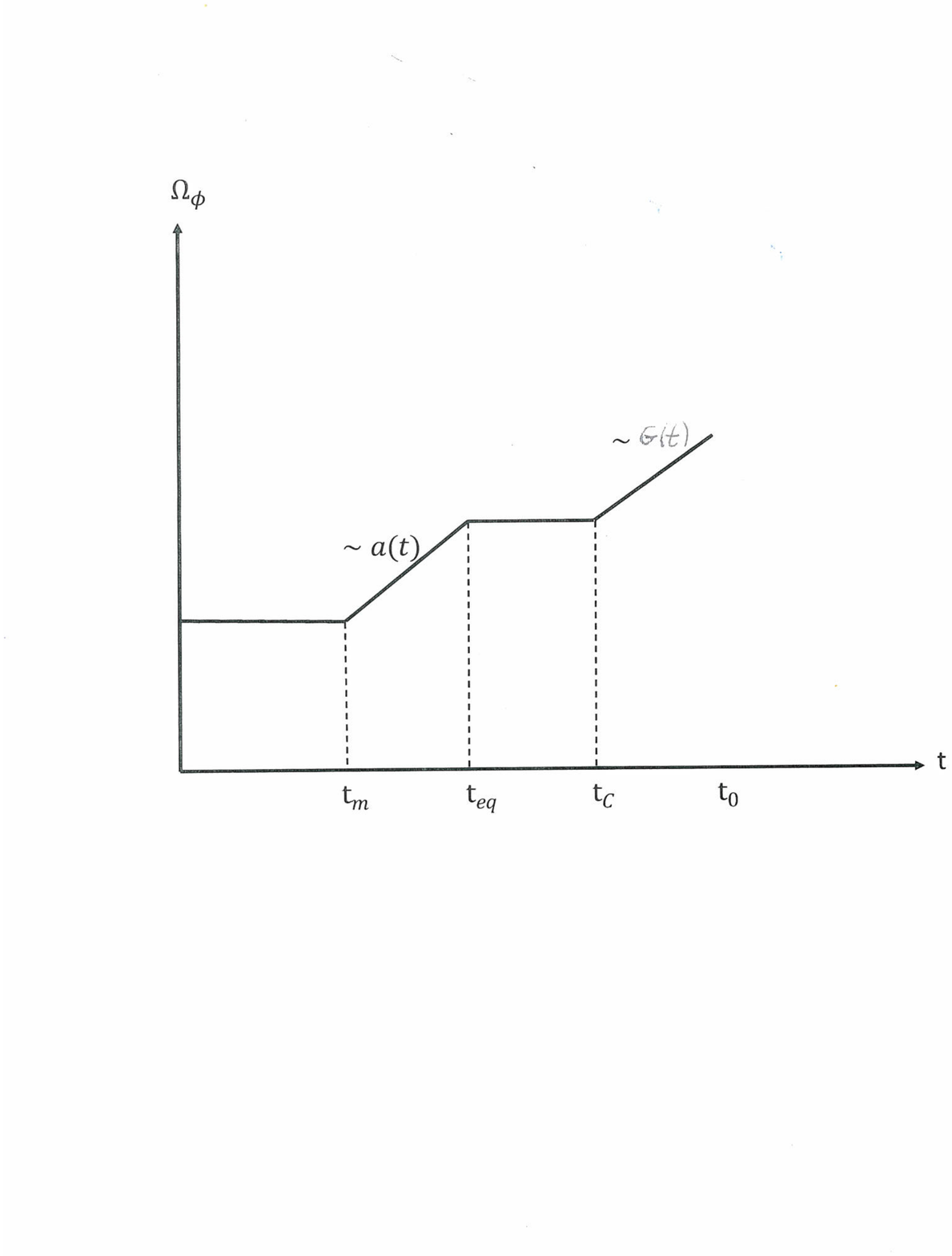}
\caption{Sketch of the time evolution of the fractional contribution $\Omega_{\phi}$
of the $\phi$ field to energy density of the Universe. The horizontal axis is time,
the vertical axis is the value of $\Omega_{\phi}$. The contribution is constant until
the time $t_m$ when the gauge field mass becomes important. It then rises as
the scale factor, to become constant again for $t > t_{eq}$. Once the secular
growth of the $E$ field becomes dominant at the time $t_c$ the contribution of
$\phi$ to $\Omega$ once again begins to rise as given by the secular growth
factor $G(t)$. Note that $t_0$ is the present time.}
\end{center}
\end{figure*}

Since the energy density of the new gauge field is larger than 
${\rm tr} ({\vec E} \cdot {\vec B})$, 
(by the Schwarz inequality), a necessary condition for $\phi$ to be a
dark energy candidate is that 
\be
V(\phi) \, \gg \, \text{tr }({\vec E} \cdot {\vec B}) \, ,
\ee
which, by (\ref{SReq2}), can only hold provided
\be
\lambda \, \gg \, 1 \, .
\ee

For $\phi$ to be a good {\it tracking quintessence} candidate, we
need the secular growth term in (\ref{growth}) to become dominant
at a time $t_c$, with
\be
t_{eq} < t_c < t_0 \, ,
\ee
where $t_0$ is the present time. This leads to the requirement
that $B(t_i) / E(t_i)$ needs to be slightly smaller than $\lambda^{-1}$,
which is a tuning condition we need to impose.

Above we have assumed that the slow-roll conditions
\ba
{\ddot \phi} \ll V'(\phi) \quad \text{and  } \text{   }
3 H {\dot \phi}  \ll V'(\phi)
\ea
are satisfied, and that the equation of state of $\phi$
leads to acceleration. It is easy to check
that the slow-roll conditions are
satisfied, provided
\be \label{SRcond}
f \, \ll \, m_{pl} \, .
\ee
It is not hard to check that the equation of state for $\phi$ is dominated
by the potential energy if condition (\ref{SRcond}) is satisfied. Thus,
the field $\phi$ is indeed a candidate for tracking dark energy.

Finally, we study the magnitude of the contribution of $\phi$ to
the dark energy budget. Evaluating (\ref{tracking}) at the present
time $t_0$ and assuming $t_0 > t_c$ we obtain (dropping Lie-algebra indices on $\vec{E}$ and $\vec{B}$)
\be \label{Omega}
\Omega_{\phi}(t_0) \,  \simeq  \, \frac{\lambda}{8} \frac{({\vec{E}} \cdot {\vec{B}})(t_i)}{\rho_r(t_i)} 
\bigl( \frac{a(t_{eq})}{a(t_m)} \bigr) \lambda \frac{B(t_i)}{E(t_i)} {\rm log}(\frac{t_0}{t_i}) \, .
\ee
If we do not want to introduce a new mass hierarchy into our model, it
is natural to assume that $t_m \sim t_i$. In this case, an initial value of
$\text{tr }({\vec{E}} \cdot {\vec{B}})$ comparable to the initial matter density is
required in order for the order of magnitude of (\ref{Omega}) today
to be close to unity. This is ensured 
if, initially, at time $t_i$, the ratio between the instanton density and the energy density of radiation is 
proportional to the ratio between baryon- and entropy density, i.e., 

\be
\frac{\text{tr }({\vec{E}} \cdot {\vec{B}})(t_i)}{\rho_r(t_i)} \, \sim \, \frac{n_B}{s}(t_i), \,
\ee
where $n_B$ is the baryon number density and $s$ the entropy density. Hence,
the smallness of the initial contribution of $\phi$ to dark energy is guaranteed by
the observed small baryon to entropy ratio. This factor is believed to be
of the order $10^{-10}$.

\section{Particle Physics Connections}

{\bf A) An axion coupling to an anomalous matter current}:
\noindent

The standard axion field, $a(x, t)$, along with the Peccei-Quinn symmetry 
has been introduced to solve the strong CP problem of QCD; see \cite{PQ}.
The mechanism leading to the spontaneous breaking of the Peccei-Quinn symmetry involves
a complex scalar field with a standard symmetry breaking
potential whose angular variable is the axion field $a$ \cite{axion}. 
The coefficient of the Pontryagin density, ${\rm{tr}} (F \wedge F)$, in the QCD Lagrangian
then becomes a dynamical variable.  At the perturbative
level, the axion has a flat potential. Non-perturbative instanton
effects create however a non-trivial potential, $V(a)$, for the axion. This
potential is periodic in $a$, which is an ``angular variable.''
The periodicity of the potential is not unproblematic, since it could
give rise to an axion domain-wall problem. 

The axion of QCD is a candidate for dark matter \cite{axionDM}, but cannot be
a candidate for dark energy, since it interacts too strongly
with electromagnetism. Any viable candidate scalar field
for dark energy needs to couple very weakly to standard
model matter \cite{Carroll}. 

The idea underlying our proposal is that the field $\phi$ responsible for 
dark energy could be a new axion field conjugate to an anomalous matter current \cite{ABJ}. 
A possible example would be an anomalous lepton current, in which case the
gauge field would be a weak $SU(2)$ field (see, e.g.,  \cite{Wein} ). 
The gradient of $\phi$ can then be linearly coupled to the anomalous 
axial vector current $J_5^{\mu}$, introducing a term proportional to
\begin{equation} \label{couplingterm}
\partial_{\mu}\phi \cdot J_{5}^{\mu}
\end{equation}
in the Lagrangian of the theory. Apparently, the time derivative of $\phi$ then plays the role 
of a space-time dependent axial chemical potential for an axial charge density \cite{PF}, (e.g.,
the $0$-component of the left-handed lepton current). 
This may furnish an ingredient in a mechanism responsible for the observed matter-antimatter asymmetry. 
Thanks to the anomaly equation \cite{ABJ}, the term (\ref{couplingterm}) 
is equivalent to a term proportional to
\begin{equation} \label{fiftytwo}
\phi (F \wedge F + \frac{1}{f} \sum_{j} m_{j} {\bar{\psi}}_{j} \gamma_{5} \psi_{j}) \, ,
\end{equation}
with $j$ labeling fermion species, where species $j$ has mass $m_{j}$ and is described by a spinor field $\psi_{j}$.

Instanton effects are usually expected to lead to a potential for $\phi$ that is
periodic in $\phi$, and this possibility is studied in forthcoming work.
One may imagine, however, that axion shift-symmetry
breaking effects might generate an exponential potential. 
The value of the parameter $f$ is related to the symmetry
breaking scale, and the energy-scale parameter $\mu$ is set 
by the strength of the instanton effects. 

\noindent{}

{\bf B) Universal axion of string theory}:
\noindent

Axions arise naturally in superstring theory \cite{Witten}. Specifically,
string compactifications generate Peccei-Quinn type
symmetries often broken at the string scale \cite{Witten2}.
For example \cite{Gukov}, there is an axion field $a$ that is in the same 
chiral superfield $S$ as the four dimensional dilaton $\varphi$
\be
S \, = \, e^{- \varphi} \, + \, i a \, .
\ee
In addition, there is an axion field $\tilde{a}$ in
the superfield ${\tilde S}$ of the volume scalar $\rho$:
\be
{\tilde S} \, = \, e^{\rho} \, + \, i {\tilde a} \, .
\ee

The Peccei-Quinn symmetries of string theory are
always broken by stringy instanton effects, leading to
a coupling of the axion to some ${\rm{tr}} (F \wedge F)$- term.
This can be shown explicitly by reducing the ten-dimensional
supergravity action to four space-time dimensions via
compactification on some internal Calabi-Yau manifold;
(see e.g. \cite{Gukov}). Such a compactification also generates
potentials for the superfields to which the axions belong.
These potentials are typically exponential in the radial direction,
but a remnant of the exponential potential may also
affect the potential in the axion direction; especially if
stringy effects lead to a breaking of the shift symmetry
in the axion direction, as happens in axion-monodromy models
\cite{Silverstein, Westphal}. For some explicit constructions of
exponential potentials see \cite{Stephon2}.

\noindent{}

{\bf C) Axion monodromy}:
\noindent

Indeed, it has recently been realized that stringy effects break the shift symmetry
of the axion. The axion ceases to be an angular variable and,
instead, has an infinite range of values. Monodromy induces an axion
potential rising without bound, as $\phi$ increases to $\infty$; see, e.g., \cite{Westphal}. 
At large field values, the axion potential may be linear. To make
contact with our scenario we need to assume that the potential
is exponential at small field values.

We are not the first to connect an axion with a potential induced
by stringy monodromy effects with dark energy. In \cite{Trivedi}
it was in fact suggested that a stringy axion may play the role of a 
quintessence field. The construction in \cite{Trivedi} makes use of
standard slow-roll inflation and thus requires super-Planckian
field values. It must still be shown that such field values 
are consistent from the point of view of
string theory, since for other axion models they are not \cite{Rudelius}. 
In our construction, the axion field
values are sub-Planckian, because slow-rolling is induced by
the coupling of the axion to the Chern-Simons term of a gauge field.

\section{Conclusions and Discussion}

We have studied a model of {\it tracking dark energy} in which
dark energy arises from an axion field $\phi$ linearly coupled to the Pontryagin density
of a gauge field, i.e., to a term ${\rm{tr}} (F \wedge F)$. Thanks to this coupling,
 the axion is rolling slowly even for sub-Planckian
field values, It thus has the right equation of state to account
for dark energy. We have considered the example of an exponential
potential for the axion. The coupling
between the axion and the gauge fields leads to a secular growth term 
in the electric field. At early times,
the energy density in $\phi$ tracks that of the background matter;
but when the secular growth term becomes important the 
contribution of $\phi$ to the density parameter $\Omega$ starts to increase.
We have studied the evolution of $\Omega_{\phi}$ (the fraction
of the total energy density required for a spatially flat universe due to the axion field $\phi$)
as a function of time and found that it is constant for early times 
$t < t_m$, where $t_m$
is the time when the mass of the gauge field becomes important. It
grows linearly in the scale factor between time $t_m$ and  the time, $t_{eq}$, 
of equal matter and radiation. After time $t_{eq}$, the value of  $\Omega_{\phi}$
ceases to grow until the time when the secular growth term becomes dominant,
after which it will start to grow again.

In order for $\phi$ to be a successful candidate for dark energy,
the time when $\Omega_{\phi}$ approaches $\Omega = 1$ has
to be close to the present time $t_0$. This is only the case if, at the initial time,
the $\phi$- field energy is a small contribution to the total energy density.
Our proposal is that this small initial value of $\Omega_{\phi}$
is linked to the small value of the lepton to entropy ratio. This would imply
that the secular growth term becomes important only at rather
late times. Thus, our model represents an implementation of
the ``tracking quintessence'' scenario of \cite{tracking}.
We have shown that, in order to obtain the currently
observed value of dark energy in our model, it suffices to require 
a fairly mild tuning of dimensionless coupling constants.

In this paper we have neglected the coupling of the axion field $\phi$ 
in (\ref{fiftytwo}) to pseudo-scalar mass terms,
$m_{j} {\bar{\psi}}_{j} \gamma_{5} \psi_{j}$, of fermionic matter fields.
Taking such couplings into account would lead to extra terms on the right side of the axion
equation of motion (\ref{axionEq}). For $H > \text{max}_{j}m_j$, these terms will decay as 
radiation, and, for $H < \text{min}_{j} m_j$, they decay
as matter. If $\bar{m} \equiv \text{max}_{j} m_{j} < m$, there is a time interval
$t_m < t < t_{\bar{m}}$ when the contribution due to the mass terms 
on the right side of (\ref{couplingterm}) decays rapidly, relative to the one of the $F \wedge F$ term. 
Hence, as long as $m > \bar{m}$, the extra terms in (\ref{couplingterm}) will not change our
conclusions.

It has been pointed out that if the field responsible for dark energy is a pseudo-scalar 
field then it could couple to visible matter, and this leads to rather stringent
constraints.  A discussion of the coupling of an axion to visible matter
has been given in \cite{Carroll}, where it has been assumed that the axion 
couples to the $\vec{E}\cdot \vec{B}$- term of electromagnetism. This would lead to 
a rotation of the direction of polarization of light emitted by distant radio sources.  
The constraints resulting from this effect are quite restrictive and would potentially 
rule out our model if our axion were to couple to the electromagnetic field.  
However, we have assumed that our axion does \textit{not} interact with the photon 
and thus evades the bounds presented in \cite{Carroll} and in related work.  
In a future paper, we will investigate collider signals due to a possible coupling of the 
axion field $\phi$ to W- and Z- bosons.

\section*{Acknowledgement}
\noindent

One of us (RB) wishes to thank the Institute for Theoretical Studies of the ETH
Z\"urich for kind hospitality. RB acknowledges financial support from Dr. Max
R\"ossler, the Walter Haefner and ETH Zurich Foundations, and
from a Simons Foundation fellowship. The research of RB is also supported in
part by funds from NSERC and the Canada Research Chair program.
JF thanks A. H. Chamseddine and D. Wyler for very helpful discussions on related
ideas. SA thanks JiJi Fan and Sam Cormack for helpful discussions, and
acknowledges support from the US Department of Energy under grant
DE-SC0010386. We wish to thank Tom Rudelius and Gary Shiu for
insightful communications.

\end{document}